\documentclass[]{aa}  
\usepackage{graphicx}
\usepackage{txfonts}
\usepackage{hyperref}
\usepackage{rotating}
\usepackage{natbib}
\usepackage[usenames]{color}
\usepackage{multirow}
\usepackage{url}
\usepackage{ulem}
\usepackage{booktabs} 
\usepackage{tabulary} 
\usepackage{siunitx}
\bibpunct{(}{)}{;}{a}{}{,}

\newcommand{\kms}   {{\rm \  km \  s^{-1}}}
\newcommand{\K}   {{\rm \,K}}

\newcommand{\micron}{\mbox{$\mu$m}}

\newcommand{\msun}{M$_{\odot}$}   
   
\newcommand{\CFC}{diffuse line of sight cold cloud}
\newcommand{\AF}{absorbed fraction}   

\newcommand{\hoA}{o-H$_2^{18}$O 1$_{10}$-1$_{01}$}  
\newcommand{\hoC}{o-H$_2$O 1$_{10}$-1$_{01}$}       
\newcommand{\hoE}{p-H$_2$O 2$_{02}$-1$_{11}$}       
\newcommand{\hoI}{p-H$_2^{18}$O 1$_{11}$-0$_{00}$}  
\newcommand{\hoK}{p-H$_2$O 1$_{11}$-0$_{00}$}       
\newcommand{\hoN}{o-H$_2$O 2$_{12}$-1$_{01}$}       
\newcommand{\CO}{$^{13}$CO\ (10-9)}                 
\newcommand{\COa}{C$^{18}$O\ (9-8)}                 

\def\la{\lower.5ex\hbox{$\; \buildrel < \over \sim \;$}}
\def\ga{\lower.5ex\hbox{$\; \buildrel > \over \sim \;$}}

\begin{document} 

\title{Structure and kinematics of the clouds surrounding \\the Galactic mini-starburst W43 MM1} 
\titlerunning{Structure and kinematics of the clouds surrounding W43 MM1}

\author{T. Jacq\inst{1} \and J. Braine\inst{1}  \and F. Herpin \inst{1}  \and F. van der Tak \inst{2,4} 
  \and F. Wyrowski \inst{3}  }

  \institute{Laboratoire d'astrophysique de Bordeaux, Univ. Bordeaux, CNRS, B18N, allée Geoffroy Saint-Hilaire, 33615 Pessac, France\\
             \email{thierry.jacq@u-bordeaux.fr}
        \and
  SRON Netherlands Institute for Space Research, PO Box 800, 9700AV, Groningen, The Netherlands
        \and
  Max-Planck-Institut f\"ur Radioastronomie, Auf dem H\"ugel 69, 53121 Bonn, Germany
        \and
  Kapteyn Astronomical Institute, University of Groningen, The Netherlands
   }
\date{Received xxxx; accepted xxxx}

\abstract { 
Massive stars have a major influence on their environment yet their formation is difficult to study as they form quickly in highly obscured regions and are rare, hence more distant than lower mass stars.  W43 is a highly luminous galactic massive star forming region at a distance of 5.5\,kpc and the MM1 part hosts a particularly massive dense core (1000\,\msun within 0.05 pc).  We present new Herschel HIFI maps of the W43 MM1 region covering the main low-energy water lines at 557, 987, and 1113\,GHz, their H$_2^{18}$O counterparts, and other lines such as \CO\ and \COa\ which trace warm gas. These water lines are, with the exception of line wings, observed in absorption.  Herschel SPIRE and JCMT $450\,\micron$ data have been used to make a model of the continuum emission at the HIFI wavelengths.  Analysis of the maps, and in particular the optical depth maps of each line and feature, shows that a velocity gradient, possibly due to rotation, is present in both the envelope ($r\ga 0.5$\,pc) and the protostellar core ($r\la 0.2$\,pc). Velocities increase in both components from SW to NE, following the general source orientation.  While the H$_2$O lines trace essentially the cool envelope, we show that the envelope cannot account for the H$_2^{18}$O absorption, which traces motions close to the protostar.  The core has rapid infall, $2.9\kms$, as manifested by the H$_2^{18}$O absorption features  which are systematically red-shifted with respect to the \CO\  emission line which also traces the inner material with the same angular resolution.  Some H$_2^{18}$O absorption is detected outside the central core and thus outside the regions expected (from a spherical model) to be above 100\K -- we attribute this to warm gas associated with the other massive dense cores in W43 MM1.
Using the maps to identify absorption from cool gas on large scales, we subtract this component to model spectra for the inner envelope.  Modeling the new, presumably corrected, spectra results in a lower water abundance, decreased from 8 $10^{-8}$ to 8 $10^{-9}$ , with no change in infall rate.
}


\maketitle

\section{Introduction}\label{Introduction}

Massive star formation is still poorly understood and, while various means have been proposed to overcome the radiation pressure problem, observational ambiguities remain \citep[see reviews by][]{Zinnecker2007, Beuther2007}.  Massive dense cores are generally fairly distant and their centers always extremely embedded, making them difficult to observe with the necessary detail.  In addition, the protostellar phase appears very short \citep[a few $10^5$ years, ][]{2014prpl.conf..149T, Beuther2007}, further reducing the number of objects. A main question is how to "feed" a massive protostar, particularly in the earliest phases when it is weak in the infrared.  

Understanding high-mass star formation is one of the goals of the Water In Star forming regions with {Herschel\footnote{Herschel is an ESA space observatory with science instruments provided by European-led Principal Investigator consortia and with important participation from NASA.}} program \citep[WISH,][]{vandishoeck2011}.  Water is a potentially very abundant molecule but the vast majority of water is found as ice, {which cannot be used to study kinematics}.  Only when the temperature reaches 100\K\ is the water ice on dust grains vaporized, provoking a radical change in water vapor abundance from some $10^{-8}$ to $1-10 \times 10^{-5}$ \citep{vdtak2010,Herpin-W43-2012}.  This jump in abundance makes it possible in principle to use the water molecule to focus on the central regions.  High-energy transitions of other molecules also come from the dense and warm central regions.  However, low-energy water transitions can be observed in both emission and absorption coming from the central regions, allowing potentially infalling cooler material to be detected in absorption, unlike in high-energy transitions  \citep{2013A&A...554A..83V}.  

This idealized picture does not hold for the main water isotope as even the low-abundance gas is highly optically thick.  However, for the H$_2^{18}$O and H$_2^{17}$O isotopomers with typical abundances of respectively 450 and 2000 times lower than the main isotope, the above picture should be more appropriate. 
The goal of this paper is to {map W43 MM1 surroundings with} a large set of lines to develop a more precise image of the amount of turbulence, infall rate, and rotation in the central core but also in the cooler surrounding material.  As part of this, we try to identify the lines that best allow us to study the inner and outer parts of the protostellar envelope.

W43 MM1 is a high-mass proto-stellar object situated at about 5.5\,kpc from the sun \citep{2014ApJ...781...89Z}  at the near end of the Galactic bar.  W43 is sometimes considered to be a "mini-starburst" region because of the currently high level of star formation occurring there \citep{Motte2003}, probably due to a cloud collision \citep{Luong2013}.  Herschel observations, including water lines, towards the center of W43 MM1 were analyzed by  \citet{Herpin-W43-2012}, who present a spherically symmetric model for the structure of W43 MM1 showing that both accretion and radially increasing turbulent velocities are required.  They estimate the water abundances to be $8 \times 10^{-8}$ where $T_{mb}<100$\K\ and $1.4 \times 10^{-4}$ in the warm regions, higher than in most cores \citep{Chavarria2010}.  The water abundance in cold outer envelopes is usually lower, of order $10^{-9}$ --  $10^{-10}$ \citep{vandishoeck2013}.  

The general source structure is shown in Figure 1 of \citet{Louvet-2014-w43mm1}: a double central core (N1a and N1b) surrounded by an elongated structure with a column density of $\sim 10^{23}$ cm$^{-2}$.   \citet{Louvet-2014-w43mm1} found a number of Massive Dense Cores (MDC) in W43 MM1, mostly aligned to the NE and SW of the main core with a typical separation of about 5 to $10\,\arcsec$.  The maps presented in this work cover the region with the cluster of MDCs identified by  \citet[][Figure 1]{Louvet-2014-w43mm1}.  The main MDC (N1) dominates the cluster, with an estimated mass of 2100\,\msun, with the other MDC masses decreasing  to typically 100\,\msun~away from the center of the cluster.  All of the MDCs identified by \citet{Louvet-2014-w43mm1} in W43 appear overdense compared to other dense cores, such that they are likely to contain protostellar objects.  Thus, the central spherically symmetric model may not well represent the source structure given the presence of heating sources -- the MDCs -- distributed over roughly an arcminute, particularly to the South West of the center of W43 MM1.

As part of the WISH project, W43 MM1 was mapped with the HIFI heterodyne spectrometer aboard the Herschel Space Observatory \citep{Pilbratt, roelfsema2012} at 557, 987, and 1113\,GHz.  These frequencies correspond to the \hoC, \hoE, and \hoK\ water lines but also allow detection of the \hoA, C$^{18}$O $9-8$, \hoI, and $^{13}$CO $10-9$ lines.  The water lines (all isotopes), show a mixture of emission and absorption while the CO isotopic lines are only seen in emission.  Coupled with the spatial variations, this information is used to develop a more comprehensive model of the core and envelope structure and gas motions such as infall, turbulence, and/or rotation.  

\section{Observations}
W43 MM1 was mapped at 557, 987, and 1113\,GHz 
using the HIFI instrument \citep[][]{degraauw2010}, corresponding to the \hoC, \hoE, and \hoK\ lines.  The data were taken on October 27th, 2010 (obsid 1342219193), March 11th, 2011 (obsid 1342215899), and April 21st, 2011 (obsid 1342207373). The H and V polarizations were observed simultaneously using both the acousto-optical Wide-Band Spectrometer (WBS) with 1.1 MHz resolution and the digital auto-correlator High-Resolution Spectrometer (HRS) at higher spectral resolution. We used the {\em On-The-Fly} mapping mode with Nyquist sampling and a reference position (RA$_{OFF}=$18h48min27s, DEC$_{OFF}=-1^{\circ}44\arcmin28\arcsec$). The  \hoC~line was mapped (map size $0.92\arcmin\times1.74\arcmin$) simultaneously with the \hoA~line in the lower sideband while the \hoE~line was mapped ($1.27\arcmin\times1.60\arcmin$) with the C$^{18}$O $9-8$ in the same upper sideband. The third map ($1.35\arcmin\times1.68\arcmin$) combined  the \hoK,  \hoI, and $^{13}$CO $10-9$ lines. The (0,0) position of all the maps is $\alpha_{J2000}=18^h47^m47.0^s,\delta_{J2000}=-1^{\circ}54\arcmin28\arcsec$. The observations are part of the WISH GT-KP \citep[][]{vandishoeck2011}.

The three maps {\it (see Table \ref{table_transitions} )} were obtained using an {\em On-The-Fly} mapping procedure with spectra drifting along the declination axis on columns separated by a step chosen from the beam size of the main water line frequency. The pointing slightly differs between the H and the V receivers. For \hoK, the offset is of about 3\,$\arcsec $ in Right Ascension whereas the  {\em On-The-Fly} Right Ascension step is 9\,$\arcsec $. As a consequence we have two maps shifted in Right Ascension by one third of the Right Ascension step. For spectra with a low continuum level, the signal to noise ratio of the individual spectra is too low to have a good detection of the absorption features. In order to improve this signal to noise ratio, we then decided to produce an almost regular grid of averaged spectra by dividing the mapped area into cells on a regular grid. We averaged all spectra within a cell, both in intensity and position.\\

The off position has been inspected for each targeted frequency and does not show any emission. The frequencies, energy of the upper levels, system temperatures, integration times and {\it rms} noise level at a given spectral resolution for each of the lines are provided in Table \ref{table_transitions}. Calibration of the raw data into the $T_A$ scale was performed by the in-orbit system \citep[][]{roelfsema2012}; conversion to $T_{mb}$ was done using the latest beam efficiency estimate from October 2014\footnote{http://herschel.esac.esa.int/twiki/pub/Sandbox/TestHifiInfoPage/} given in Table~\ref{table_transitions} and a forward efficiency of 0.96. HIFI receivers are double sideband with a sideband ratio close to unity \citep[][]{roelfsema2012}. The flux scale accuracy is estimated to be 10\% for band 1 and 15\% for band 4$^1$. The frequency calibration accuracy is 20 kHz and 100 kHz, respectively for HRS and WBS observations. Data calibration was performed in the Herschel Interactive Processing Environment \citep[HIPE,][]{ott2010} version 13. Further analysis was done within the CLASS\footnote{http://www.iram.fr/IRAMFR/GILDAS/} package. These lines are not expected to be polarized, thus, after inspection, data from the two polarizations were averaged together.

\begin{table*}
\caption{Herschel/HIFI mapped water line transitions towards W43 MM1. Frequencies are from \citet{pearson1991}. The rms is the noise at $\delta \nu=1.1$MHz.}             
\label{table_transitions}      
\centering                          
\begin{tabular}{lcccccccc}        
\hline\hline                 
 Setting  &       Transition & Frequency & $E_u$ &      Beam &           Map size & $\eta_{\textrm{mb}}$ & $T_{\textrm{sys}}$ &  rms\\
 map (band) &                  &     [GHz] &   [K] & [\arcsec] &        [\arcmin] &                      &                [K] & [mK]\\
\hline
  1 (1a)     &             \hoA &  547.6764 &  60.5 &      37.8 & 0.92$\times$1.74 &                 0.62 &                 80 &   27\\
  1 (1a)     &             \hoC &  556.9361 &  61.0 &      37.1 &                  &                      &                    &     \\
  2 (4a)     &             \hoE &  987.9268 & 100.8 &      21.3 & 1.27$\times$1.60 &                 0.63 &                340 &   79\\
  2 (4a)     &  C$^{18}$O $9-8$ &  987.5604 & 237.0 &      21.3 &                  &                      &                    &     \\
  3 (4b)     &             \hoI & 1101.6982 &  52.9 &      19.9 &  1.35$\times$1.6 &                 0.63 &                390 &   20\\
  3 (4b)     &             \hoK & 1113.3430 &  53.4 &      19.7 &                  &                      &                    &     \\
  3 (4b)     & $^{13}$CO $10-9$ & 1101.3497 & 290.8 &      19.9 &                  &                      &                    &     \\
 \hline                                  
\end{tabular} 
\end{table*}

\begin{figure*}
	\centering 
	\includegraphics[keepaspectratio,width=18cm]{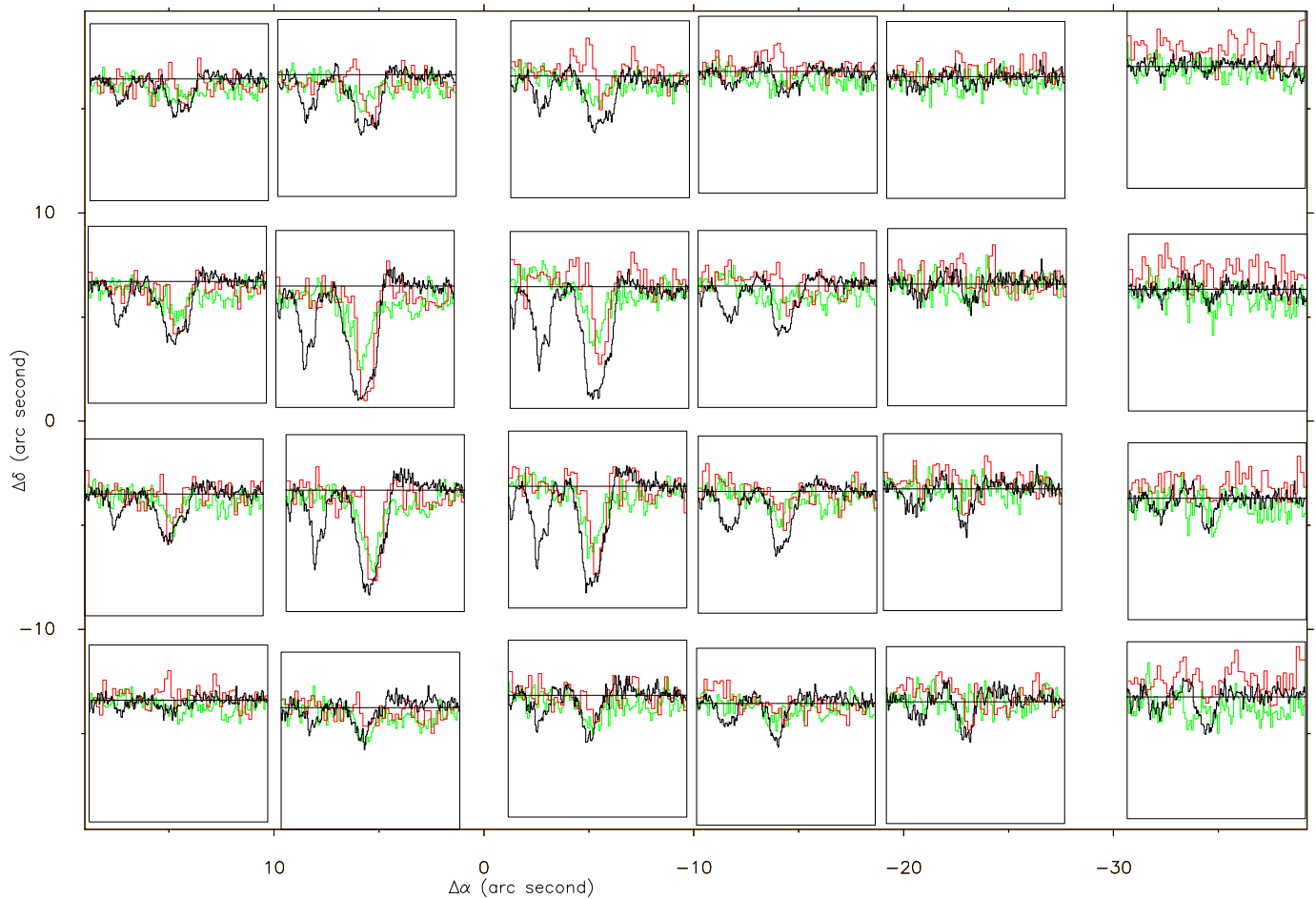} 
	\caption{{Map of observed spectra towards W43 MM1 for lines \hoK\ (black, $T_{mb}=[-2.2,\,1.00]$\,K), \hoI\ (red, $T_{mb}=[-0.55,\,0.25]$\,K), and \CO\ (green, $T_{mb}=[1.0,\,-0.50]$\,K, inverted T scale in order to ease comparisons). Velocity range: $V_{lsr}=[70,\,130]\,\kms$.} Plotted spectra are an average of OTF individual spectra around each box central position. The continuum level is subtracted.}
	\label{figs_Fig02}

\end{figure*}
\begin{figure}[!t]
	\centering
	\includegraphics[keepaspectratio,width=8.7cm]{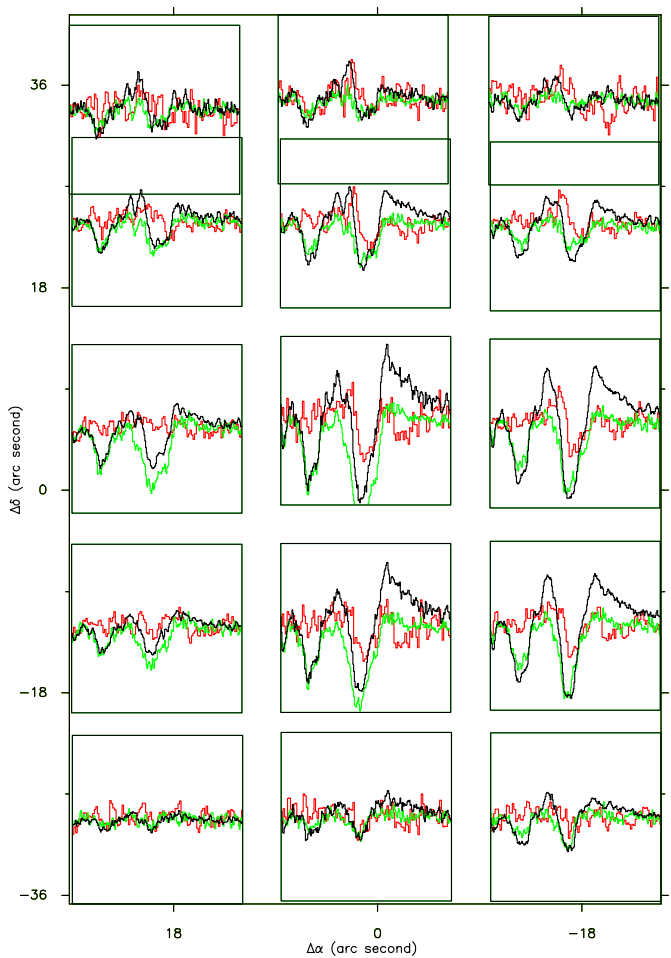}
	\caption{
	{Map of observed spectra towards W43 MM1 for lines \hoC\ (black, $T_{mb}=[-0.5,\,0.5]\K$), \hoK\ (green, $T_{mb}=[1.0,\,1.0]\K$), and \hoA\ (red, $T_{mb}=[-0.15,\,0.15]\K$). Velocity range: $V_{lsr}=[75,\,130]\kms$.} The plotted spectra are an average of OTF individual spectra around each box central position. The continuum level is subtracted. Axis units are offset ($\arcsec$) from MM1.}
	\label{figs_Fig03}
\end{figure}
\begin{figure*}
	\centering
	\includegraphics[keepaspectratio,width=18cm]{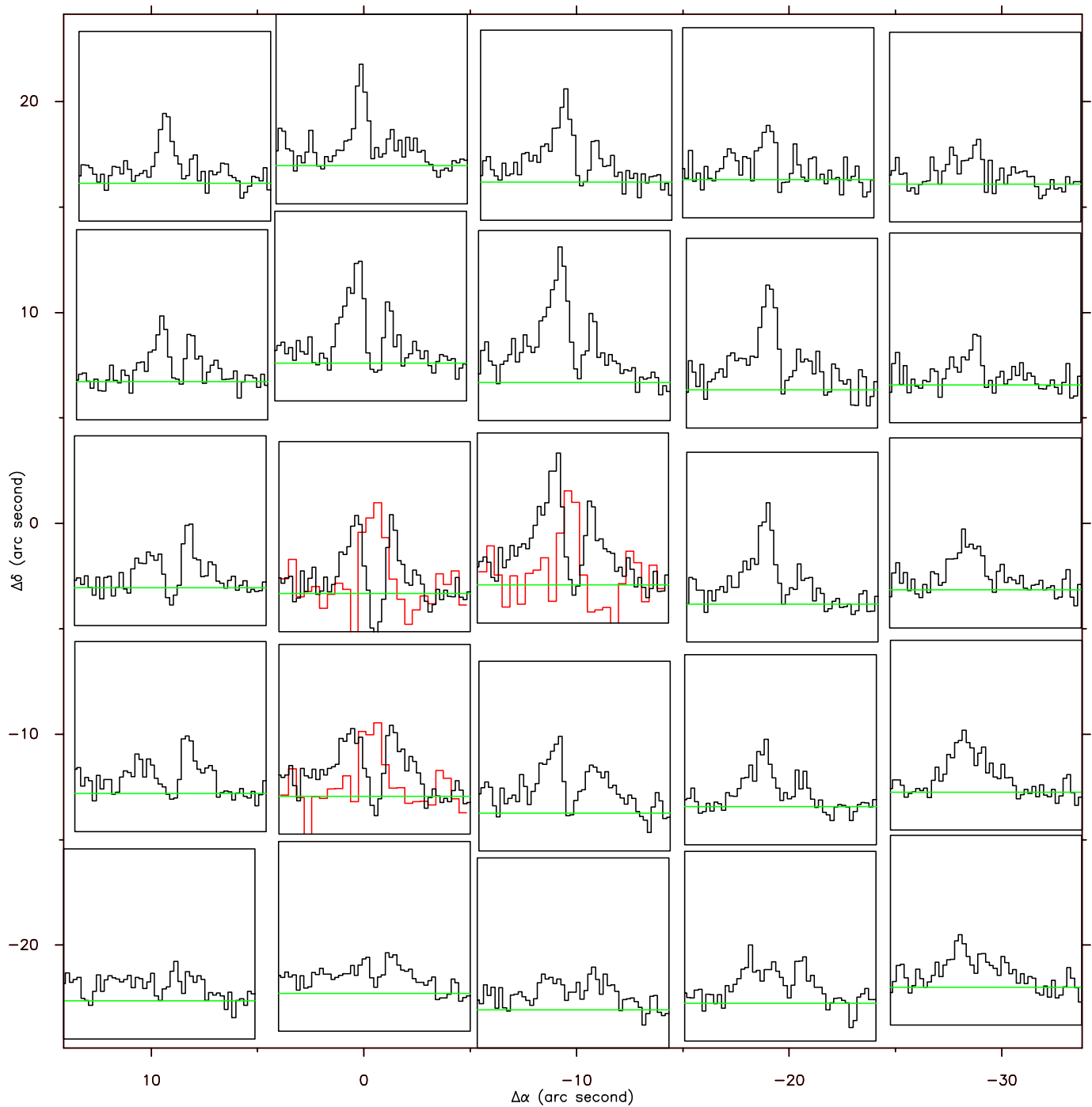}
	\caption{
	{Map of observed spectra towards W43 MM1 for lines \hoE\ (black, $T_{mb}=[-0.3,\,1.2]$\,K) and \COa\ (red, $T_{mb}=[-0.1,\,0.4]$\,K). We only plotted the three \COa\ spectra with a signal to noise ratio good enough. Velocity range: $V_{lsr}=[70,\,130]\kms$ .} Plotted spectra are an average of OTF individual spectra around each box central position. The continuum level is subtracted.  Axis units are offset ($\arcsec$) from MM1. }
	\label{figs_Fig04}
\end{figure*}
\begin{figure}
	\centering 
	\includegraphics[keepaspectratio,width=8.7cm]{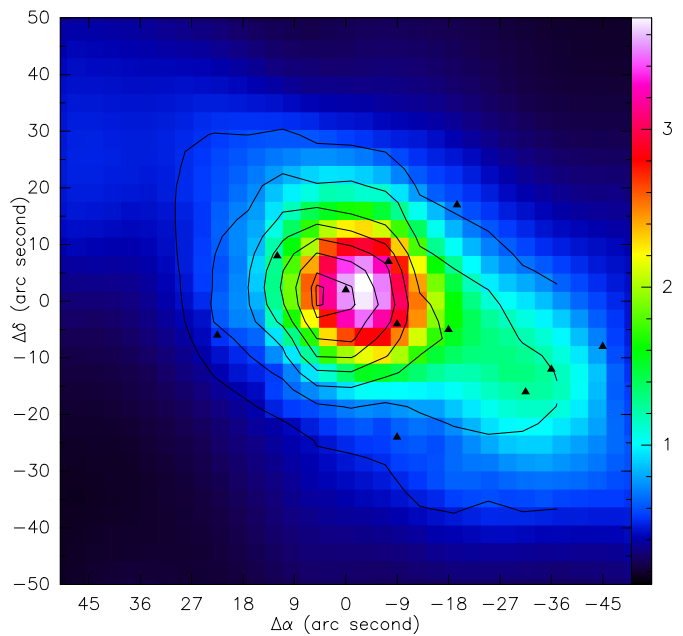} 
	\caption{Maps of the W43 MM1 continuum level at  \SI{250}{\micro\metre} (SPIRE data) and at 1113\,GHz ($\simeq $\SI{269}{\micro\metre}) before the offset correction. HIFI contour levels: 0.6 1.0 1.5 2.0 3.0 3.5 3.65 \K\ with original data positions but after continuum level correction (see text). The SPIRE data (in K) resolution is smoothed to the $20\arcsec$ beam size of \hoK. Black triangles indicate the Louvet et al 2014 sources. 
}
	 \label{figs_Fig01} 
	
\end{figure}
\begin{figure*}
	\centering
	\includegraphics[keepaspectratio,width=18cm]{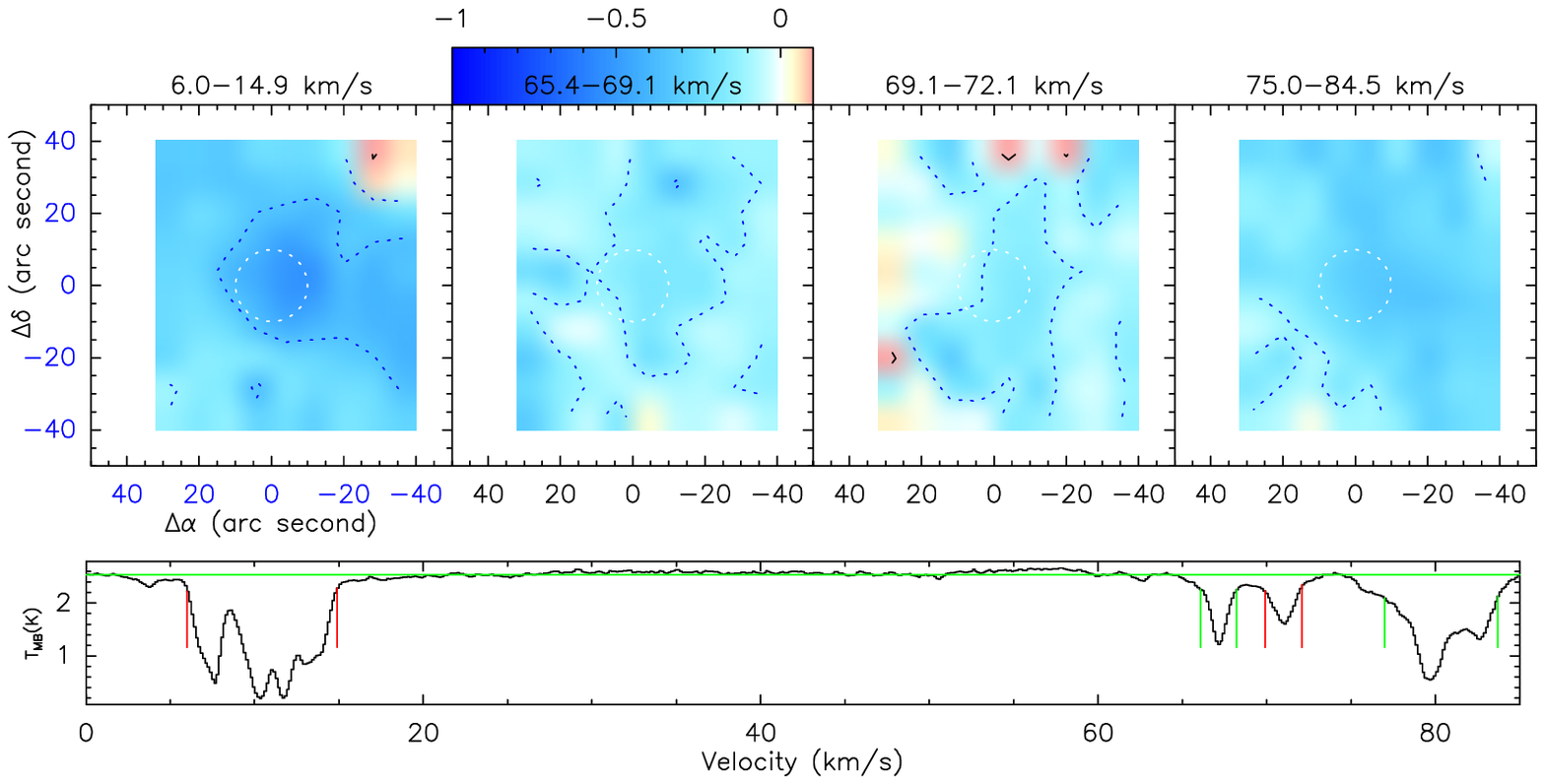}
	\caption{{Maps of the \hoK\ average absorbed fraction <$\frac{T-T_{cont}}{T_{cont}}$> in four velocity ranges bracketing the main four \CFC s. Velocity ranges are given above each box and marked on the low signal to noise spectrum observed at (0,0) by \cite{Herpin-W43-2012}, the horizontal green line marks the continuum level (K unit). The color scale is clipped in the range [-1,+0.1]. The white dashed circle stands for the beam  at 1113\,GHz at the (0,0) position. These maps show an almost constant signal over all positions having a detectable continuum level. The MM1 core, located inside the beam circle is barely visible} 
	}
	\label{figs_Fig05}
\end{figure*}
\begin{figure*}
	\centering
	\includegraphics[keepaspectratio,width=18cm,]{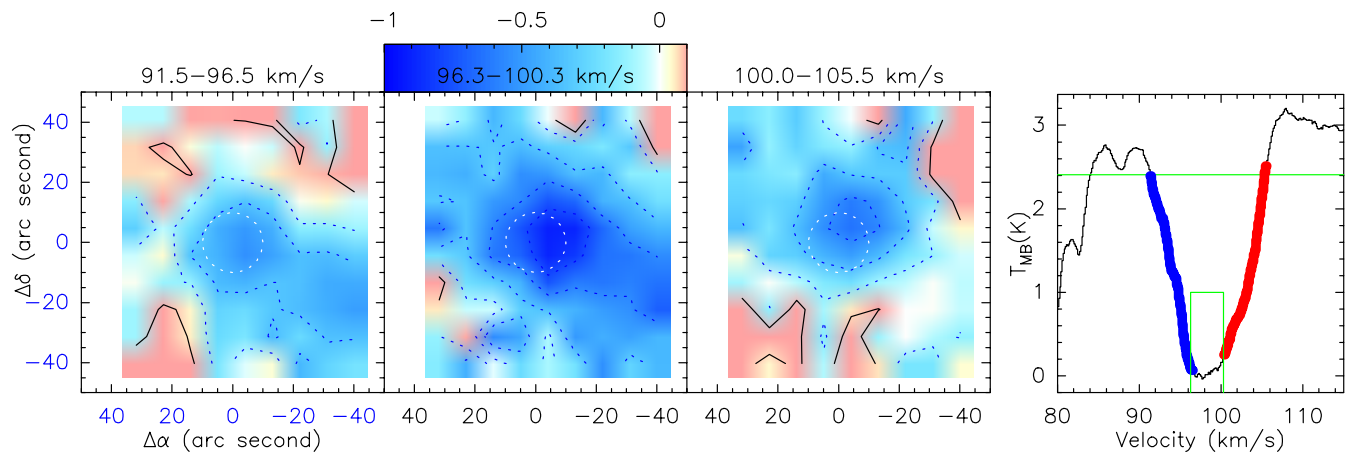}
	\caption{{From left to right: maps of the average absorbed fraction <$\frac{T-T_{cont}}{T_{cont}}$> of the main \hoK\ line blue wing, center and red wing. Right: The (0,0) spectrum as in Fig. \ref{figs_Fig05} (K units), the blue and red wing ranges are color marked on it. Other plot options are the same as in Fig. \ref{figs_Fig05}. These maps show that the blue wing material peaks south-east whereas the red wing peaks north, north-east.}
	}
	\label{figs_Fig05b}
\end{figure*}
\begin{figure*}
	\centering
	\includegraphics[keepaspectratio,width=18cm]{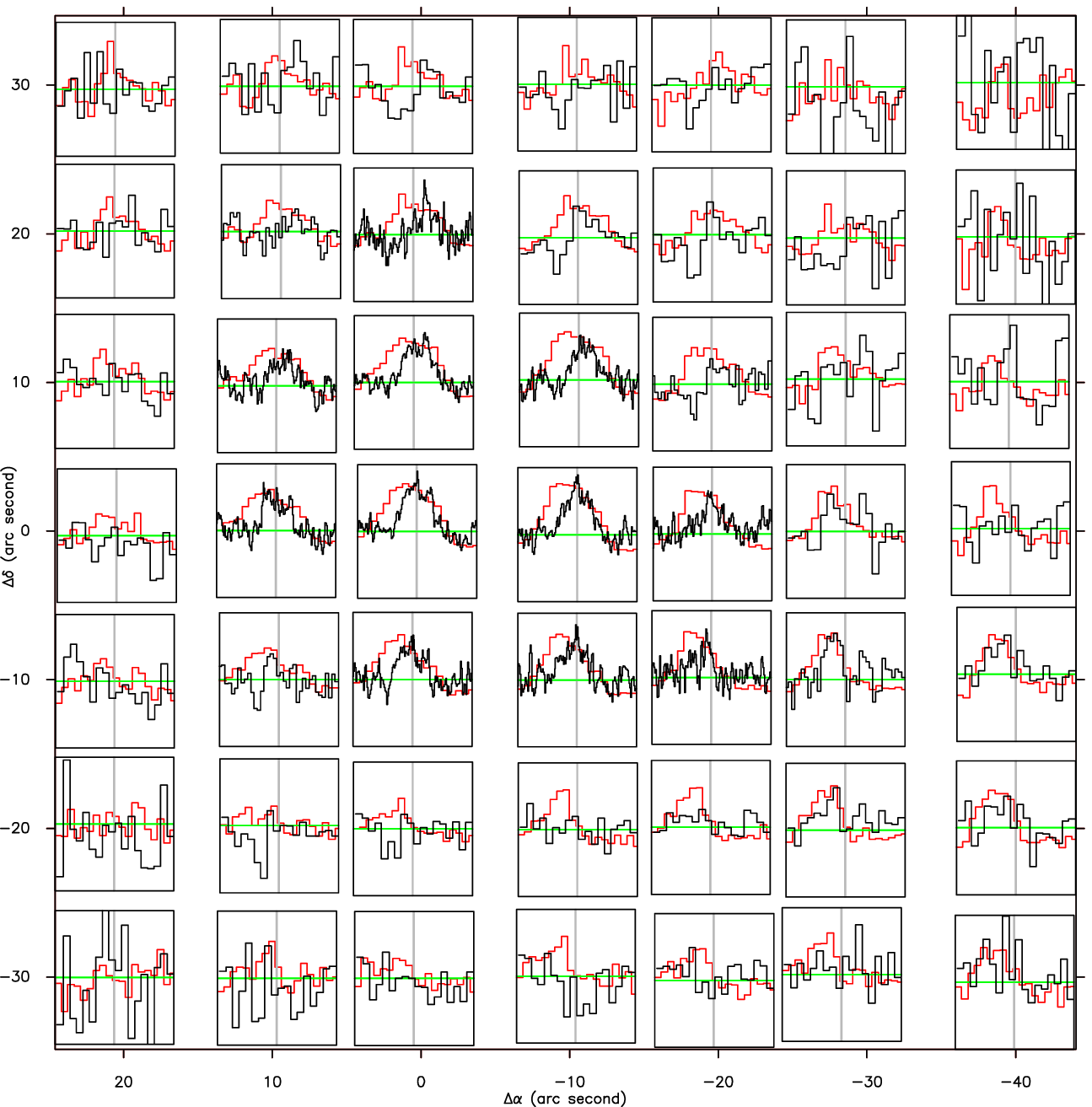}
	\caption{Map toward W43 MM1 of $\frac{T_{cont}-T}{T_{cont}}$ for \hoK\ (red) and \hoI\ (black) in the velocity range $[90\,\kms , 110\,\kms$ ]. The Y axis is from -0.1 to 1.1 for \hoK\ and, \hoI\ is scaled by 4. The horizontal green lines define the zero level. The vertical grey lines mark $100 \kms$ . {As the signal to noise ratio increases for decreasing T$_{cont}$, the spectra velocity resolution varies over the map in order to keep it at a minimum value. Hence, with the greatest T$_{cont}$, central spectra having a greater number of channels appear plotted darker than those with a lower T$_{cont}$.}  Axis units are  offset ($\arcsec$) from MM1. 
	}
	\label{figs_Fig09}
\end{figure*}
\begin{figure}
	\centering
	\includegraphics[keepaspectratio,width=8.7cm]{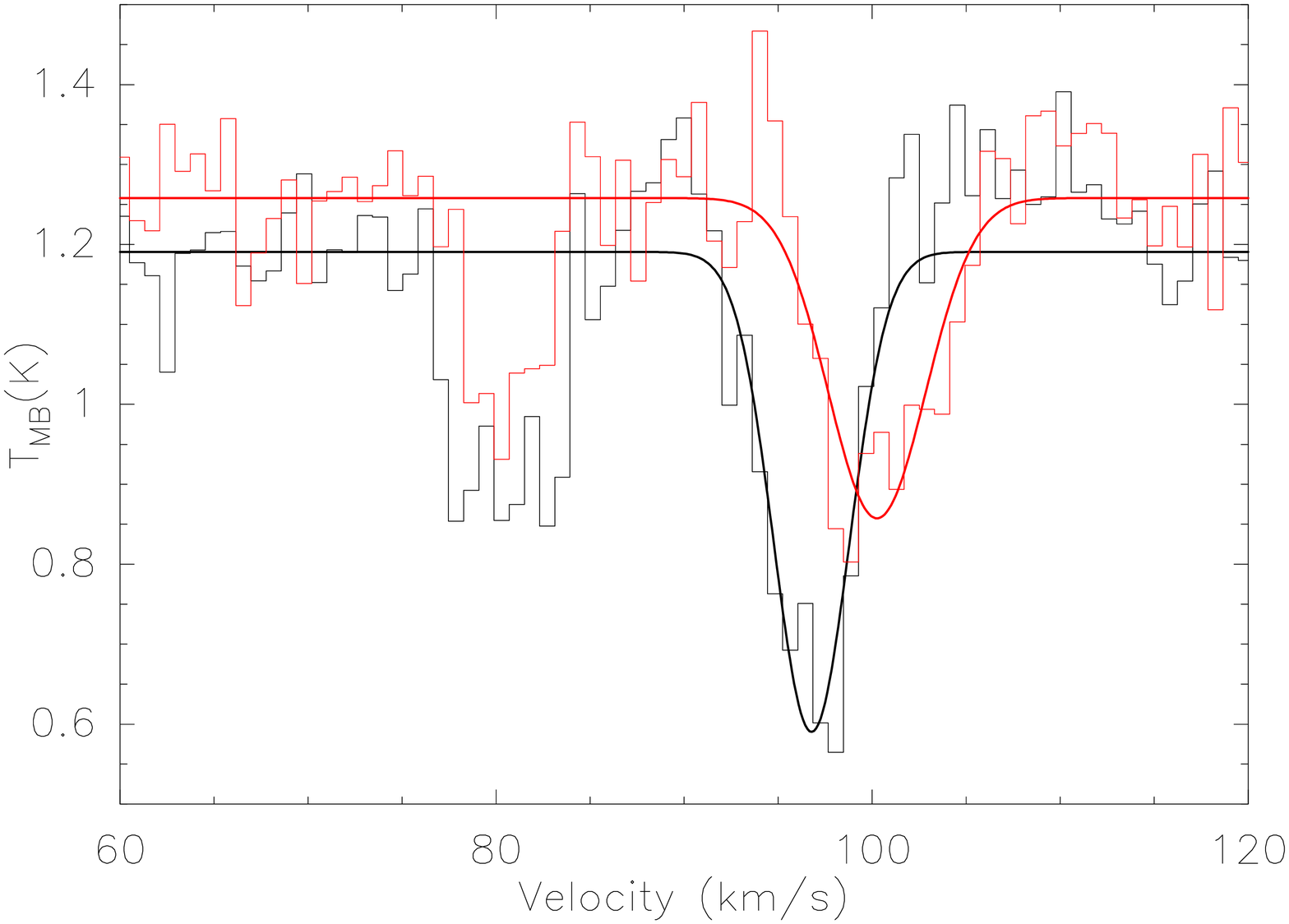}
	\caption{This plot compares \hoK\ spectra  towards our map at offsets around (-18\arcsec,-20\arcsec) in black and around (+18\arcsec,+20\arcsec) in red.
	}
	\label{figs_Fig06}
\end{figure}

\section{Mapping result} 
\label{sec:mapping_result}
The observations discussed here map, in three different frequency ranges, a region of about 80\arcsec\ by 80\arcsec\ centered on the W43 MM1 core observed, with a very good signal to noise ratio, in \citet{Herpin-W43-2012}. With a much higher rms per spectrum, our maps still exhibit the same velocity components, in varying proportions over the map: a wide ($\Delta V \simeq 40\kms$) emission, a narrower  ($\Delta V \simeq 10\kms$) emission, and a rather deep absorption of a continuum background component.
Fig. \ref{figs_Fig02}-\ref{figs_Fig04} show the spectra around MM1 for various velocity ranges bracketing the main water line and also, in Fig. \ref{figs_Fig02} and \ref{figs_Fig04}, the highest velocity \CFC\ \citep[see][]{Herpin-W43-2012}.
Strongly dominated by the MM1 core, these maps, because of an extended continuum background, still reveal structure far from MM1, making the continuum component a key feature to trace the water lines outside the MM1 core.

The analysis of  \citet{Herpin-W43-2012} needed a cold water cloud to correctly describe the various water line profiles. Unlike the core, this foreground material might be resolved in these maps 
which show that, where \hoE\ is mainly detected around MM1, the lines of \hoK\ and \hoC\ are detected at all positions having a positive continuum level. The continuum background (Fig. \ref{figs_Fig01}) which peaks at 2.5\K\ for \hoK\ at the W43 MM1 core position is almost completely absorbed for channels close to the source velocity. But both blue and red wing of the main absorption showing different behaviours around the core will help to analyze the cold foreground material close to MM1.

Fig. \ref{figs_Fig02} shows the observed spectra towards W43 MM1 for lines \hoK\ (black), \hoI\, and \CO\ (inverted to help comparison). These water spectra exhibit a strong absorption against an extended continuum. The HIFI beam does not resolve the MM1 core material, above 100\K, but is small enough to map the continuum region and hence the very close surroundings of MM1.
\\
Fig. \ref{figs_Fig03} shows the observed spectra towards W43 MM1 for lines \hoC\ , \hoK\ and \hoA. The water spectra are seen in emission and are thus mainly seen toward the MM1 core.  
\\
Fig. \ref{figs_Fig04} shows the observed spectra towards W43 MM1 for lines \hoE\  and \COa\ (restricted to spectra with a signal to noise ratio good enough). The angular resolution is similar to the map size and help very little to map the MM1 surrounding material.
\\

%
\subsection{Continuum background} 
\label{sub:continuum_background}
                                                                                 
Though our spectra do contain continuum, its precise level is somewhat uncertain. We mapped the water lines in W43 MM1 using a fast  {\em On-The-Fly} mapping HIFI-AOT which allowed to save observing time but at some price. The continuum level variation between two consecutive spectra along the drifts exhibits variations much above these expected from the spectra rms. These variations are especially large at the edges of the region where the continuum level is weaker. One even detects lines in absorption though the continuum level appears negative in the spectrum. Mapping the continuum derived from our maps produces various small artifacts due to these random level variations. The overall structure of the region is correctly reproduced but  absolute values of the continuum are unreliable at the smaller scales. As we want to use the continuum level to evaluate the opacity of the absorbing water material, it will produce large errors in parts of the map of low continuum intensity.

To overcome this problem we decided to get an estimate of the true value of the continuum level by using data obtained with SPIRE at \SI{250}{\micro\metre} 
\citep{Luong2013}. Our best map for comparison is the \hoK\ one whose wavelength is   \SI{269}{\micro\metre} quite close to the SPIRE data.
Fig. \ref{figs_Fig01} shows an image of the continuum level observed by SPIRE at  \SI{250}{\micro\metre} whose angular resolution was smoothed to the \hoK\ one.
The continuum derived from the \hoK\ spectra observed at 1113\,GHz is overplotted as contour levels.
The SPIRE map of the  continuum has a strong and unresolved peak located at the MM1 position but, at lower levels it follows an underlying aligned north-east to south-west structure anchored to the MM1 position. It decreases rapidly towards the edges of the map along the south-west map axis and more slowly towards the south-east part of the map. The smallest beam size ($\simeq$20\arcsec, \hoK ) does not resolve the 10\arcsec central MM1 source but does resolve the extended continuum structure.  Thus, having observed absorption lines more than a beam away from MM1 we can get an insight to the material surrounding the central core. 
Despite level uncertainties in the HIFI map, a comparison of both maps shows a small offset $(-4\arcsec,+2\arcsec)$ between the two peak locations. The reason of this discrepancy is unknown.   
A map of the region obtained by SCUBA/JCMT at \SI{450}{\micro\metre}  (Chavarria, priv. comm.) is also available. The SPIRE and SCUBA maps have a highly similar geometry apart from a less than $3\arcsec$ north-south shift between the peaks positions. 
Part of the data analysis which follows uses, for \hoK, the computation of the \AF\ $\frac{T-T_{cont}}{T_{cont}}$. The error on this quantity is highly dependent on  the $T_{cont}$ value. The low reliability on the \hoK\ continuum level makes it too difficult to use. Thus, we decided to modify the observed continuum level in all spectra using a model derived from the SPIRE map.

\subsection{Model of the continuum background} 
\label{sub:model_continuum_background}

We build a new set of \hoK\ spectra by setting the continuum level in each spectrum to a new value derived using the following procedure. We build the continuum model of the region at 1113\,GHz from the SPIRE map smoothed at the HIFI \hoK\ spatial resolution to get the map geometry. We then derive the intensity of the peak from the  \citet{Herpin-W43-2012} observation. For the latter, one could note that the main absorption in this spectrum goes below zero. This observation was done in a beam switch mode. The SPIRE map showed that the OFF positions were not fully free of continuum, especially the east one. Taking this into account as well as the spectral index between the USB and LSB band and also, the slight offset between this spectrum and the position of the maximum in the map we set the map maximum to 3.71\K.
The second step of our continuum building process is to shift the HIFI map by $(-4\arcsec,+2\arcsec)$ in order to correct the observed discrepancy between maps (see above). We think that this step is necessary as without it, in the final map many of the spectra having the deepest absorption will fall below the zero level. Once done, we can easily find for each spectrum the new continuum level value at the corresponding offset in the model.

\subsection{Spectral maps} 
\label{sub:spectra_lines_maps}

All data exhibit various absorption features between $0\kms$ and $140\kms$. 
At almost all map positions, even further than one \hoK\ beam size, all water lines, as well as \hoI, exhibit an often wide and well defined absorption close to the source velocity. The detection of weaker and narrower absorptions at positions distant from MM1 by much more than a beam size proves that the absorbing material extends over a region bigger than the unresolved MM1 core itself. Many narrower absorptions are also seen at lower velocities ranging from $85 \kms$ down to $0 \kms$. These were interpreted by \cite{Herpin-W43-2012} as absorbing material on the line of sight in front of the continuum background, the so-called \CFC s. This cold material is thought to be spatially unrelated to MM1 \citep[see][]{2013A&A...554A..83V}. 

The three maps discussed in this paper are of unequal interest to investigate the MM1 region structure.
\hoC\ spectra have a good rms but, an angular resolution about four times larger than the source size.
The \hoE\ map contains a few bad spectra for one of the  {\em On-The-Fly} pass, giving rise to a high signal to noise. 
\hoK, detected in absorption against the extended continuum, with the best rms among the three maps, is thus the most promising line to map the kinematics and physical structure at sizes comparable to the MM1 core. Opacity maps (Fig. \ref{figs_Fig05} and Fig. \ref{figs_Fig05b}) later derived from these data will help to separate line components over the map.

\subsubsection{The main absorption component}\label{the-absorption-component} 

At the MM1 position, the three water lines absorption velocity varies between 97.8 and $99.3 \kms$ whereas \hoC\ and \hoK\ are seen strongly absorbed at the source velocity. \hoE\ with a higher energy level is a line whose profile exhibits a well-visible emission component (see central position of Fig.\ref{figs_Fig02}-\ref{figs_Fig04}).
As one would expect, the deepest \hoK\ absorption in the map is indeed observed at the location of the strongest continuum component, i.e. the MM1 location. 
\\

\subsubsection{\hoK\ map} 
\label{sub:_hok_map}
Fig. \ref{figs_Fig02} shows spectra for \hoK , \hoI\ and \CO. The \CO\ line is plotted inverted to ease the profile comparisons. The spectral resolution differs for each line.
One sees that the \hoI\ spectra (in red) are narrower than the \hoK\ spectra and located on the red part of the \hoK\ line whereas the \CO\ line (in green) appears more blue shifted. One will also note that the \CFC s are detected only for \hoK\ and \hoC\ as already stated in \cite{Herpin-W43-2012}.

\subsubsection{\hoC\ map} 
\label{sub:_hoc_map}
Fig. \ref{figs_Fig03} shows spectra toward W43 MM1 for \hoC\ and \hoA. The \hoK\ line spectrum is added for comparison of maps.
Despite a $40\arcsec$ beam size  comparable to the map size one sees various differences between both lines over the maps.\\ 
The \CFC\  absorptions are detected in \hoC\ but not in the \hoA\ isotopic line, except for the (0\arcsec,+35\arcsec) position. The \CFC\ signal evolution correlates strongly over the map with the corresponding signal in the \hoK\ line.\\
Both the \hoC\ line (black spectrum) and the \hoA\ (red spectrum) line are well detected over all the map. 
\hoC\ appears as the merging of a deep absorption and of a wider emission component only seen for the closest spectra to MM1. The absorbing component appears narrower for \hoC\ than for the \hoK\ line (green spectrum) at the central position and much more similar on the outer part of the map. The wide emission is brighter than for \hoK.\\
The \hoA\ main absorption appears narrower than \hoC\ and is also red-shifted. 

\subsubsection{\hoE\ map} 
\label{sub:_hoe_map}
Fig. \ref{figs_Fig04} shows spectra for \hoE\ and \COa. This  {\em On-The-Fly} map was achieved in two passes. In the second one, a few group of spectra suffered from poor baselines making the resulting map inhomogeneous if using the summed spectra or with a too low rms if using only the first pass. Fig. \ref{figs_Fig04} shows the sum of the two passes. \COa\ spectra being the most impacted by the bad baseline, the central spectrum is the only one shown in Fig. \ref{figs_Fig04}.
One notes that the \COa\ line is seen in emission at the velocity of the absorbed \hoE\ component. 

Though the high rms of these spectra makes it difficult to detect too weak \hoE\ lines, we do detect these lines off the MM1 center. The $21.3\arcsec$ beam size is small enough to allow some mapping off the source core, in particular of the wings. One easily sees the line profile changes around the spectrum at the (0,0) position. The blue wing part of the emission is seen up to the (0\arcsec,20\arcsec) position whereas the red one has almost disappeared. The red wing is more pronounced south-west of MM1. There is also some reliable signal in the south-east part of the map, see the [-30\arcsec,-20\arcsec\,to\,0\arcsec] three spectra. One also notes that the red wing is still detected at the west of MM1 whereas the blue one became weaker.

\subsection{Absorbed fraction maps} 
\label{sub:absorbed_fraction_maps}

As explained above, variations in the intensities of the absorptions are dominated by the variations of the continuum background. Hence, mapping these intensities does not map water but, mostly reproduce the continuum geometrical structure over the map. However, it is possible to remove this continuum contribution by plotting the \AF\ $\frac{T-T_{cont}}{T_{cont}}$ over the mapped region. Fig. \ref{figs_Fig05} and Fig. \ref{figs_Fig05b} show the integrated value of the absorbed fraction within some selected velocity range. 
Foreground clouds detected in absorption were identified in \cite{Chavarria2010} and \cite{Herpin-W43-2012} from their velocities.  Here we use a different method based on the idea that foreground clouds are necessarily nearer and thus likely to be fairly homogeneous over the small (sub-pc) scales observed in the maps presented in this article. Fig. \ref{figs_Fig05} shows the behaviour of \AF\ over the map for the four main \CFC s. The MM1 source is within the white dashed circle. One can note that the \AF\ is almost constant at all positions following the continuum background. This favors the idea that the signal originates from some foreground material unconnected with MM1 itself.\\
Fig. \ref{figs_Fig05b}  plots \AF\ for three parts of the main \hoK\ absorption, namely the blue wing of the line, its central part, and its red wing. One notes here a different behaviour with the \CFC s. One only sees the outer part of the MM1 core, the core material itself is shielded. But, one also notes that blue and red wings behave differently as they locate symmetrically around MM1. 

Fig. \ref{figs_Fig09} compares \AF\ spectra computed for \hoK\ and \hoI. The noise in \AF\ spectra increases when $T_{cont}$ decreases making the outer part of the map too noisy. To handle this difficulty we applied a variable spectral resolution to these spectra in order to keep this noise within acceptable values. The spectral resolution decreases with $T_{cont}$, mainly with the distance to MM1. 
The continuum contribution to the geometry of the map having been subtracted, these spectra present us the spatial variation of the absorbing material. Similar \AF\ mean similar opacities. 
This figure confirms that the blue wing of \hoK\ has almost no corresponding \hoI\ counterpart on the contrary to the red one. 

\begin{figure}
	\centering
	\includegraphics[keepaspectratio,width=8.7cm]{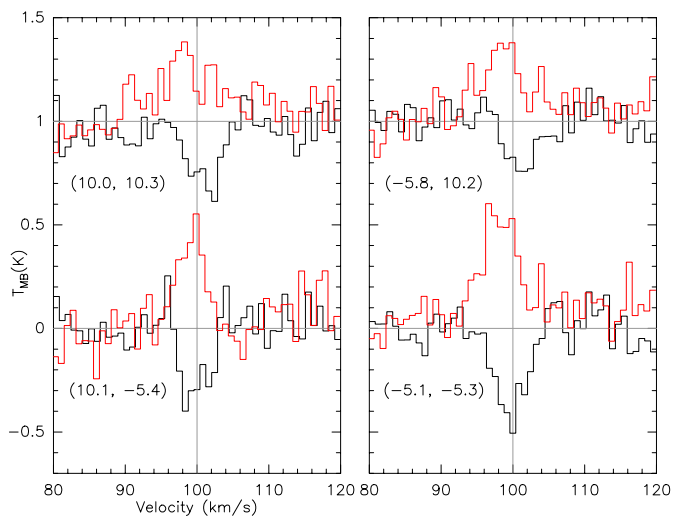}
	\caption{\hoI\ averaged spectra around the (0,0) position. The average center position is given in \arcsec\ offsets below each spectrum. Top spectra were shifted by 1K for plotting purpose. Averaged \CO\ spectra (same positions) are overplotted in red. The continuum is removed for both lines, horizontal grey lines mark the zero levels and, the vertical grey lines mark $100 \kms$.
	}
	\label{figs_Fig07}
\end{figure}
\begin{figure}
	\centering
	\includegraphics[keepaspectratio,width=8.7cm]{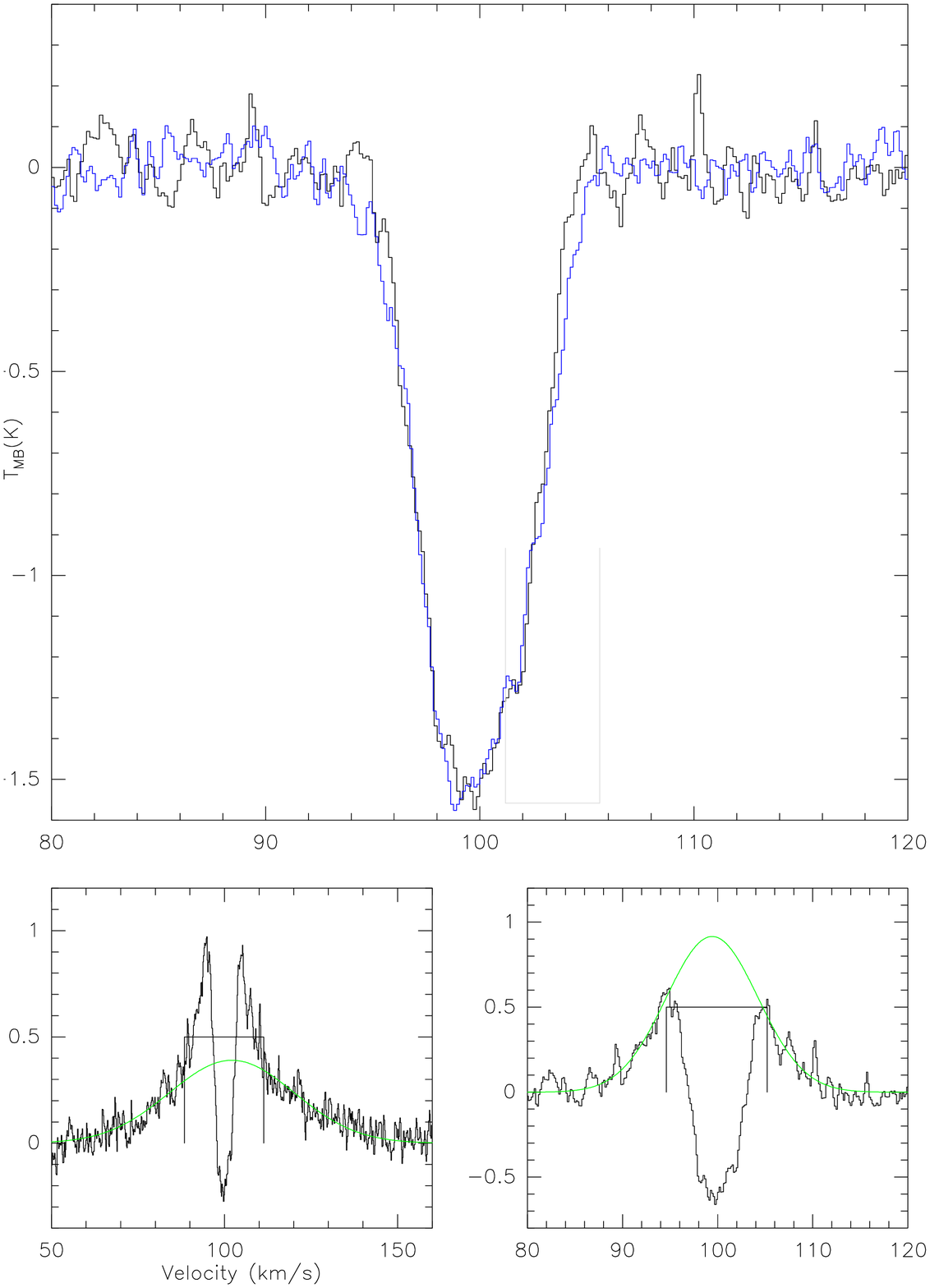}
	\caption{The \hoI\ absorption versus \hoE\ (continuum level subtracted) after a wide then narrow Gaussian components removal. The wide component is fitted in the lower left box. The wide removed spectrum is shown in the lower right box with a fit of the narrow component. The upper box shows the final result where the \hoI\ spectrum intensity is scaled to the absorption level of \hoK\ . This shows the high similarity of both profiles. Intensities are in K in all boxes.}
	\label{figs_Fig10}
\end{figure}
\begin{figure}
	\centering
	\includegraphics[keepaspectratio,width=8.7cm]{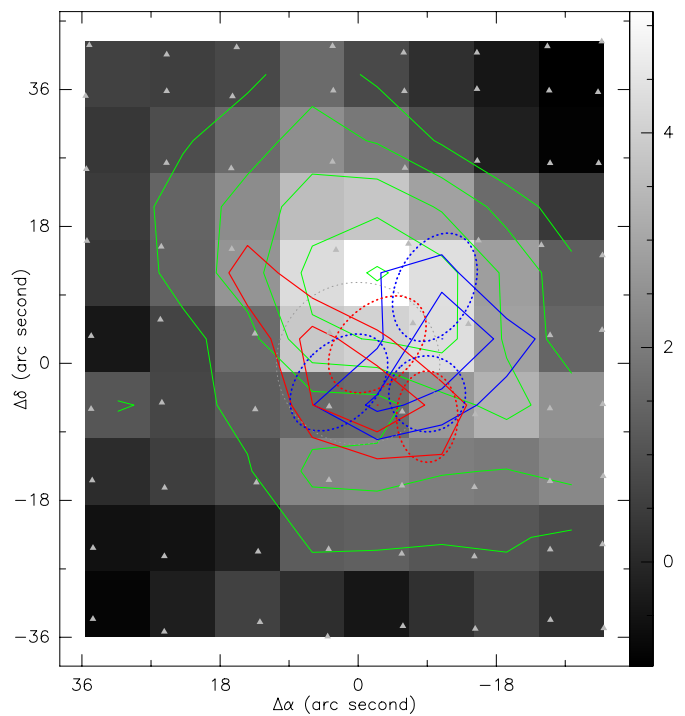}
	\caption{Map of the blue and red wings of the wide \hoE\ component. Integrated line area are in velocity ranges ($\kms$) : blue=[86,92]; red=[107,117], green=[92,107] (line center, also plotted as grey background). The grey dashed circle at the center gives the beam size. Dashed ellipses mark the position of the three CO (3-2) outflows (OF-1 and OF-2 red TBN) of Fig.2 in \cite{2014ApJ...783L..31S}. 
	}
	\label{figs_Fig08}
\end{figure}
\begin{figure*}
	\centering
	\includegraphics[keepaspectratio,width=14cm]{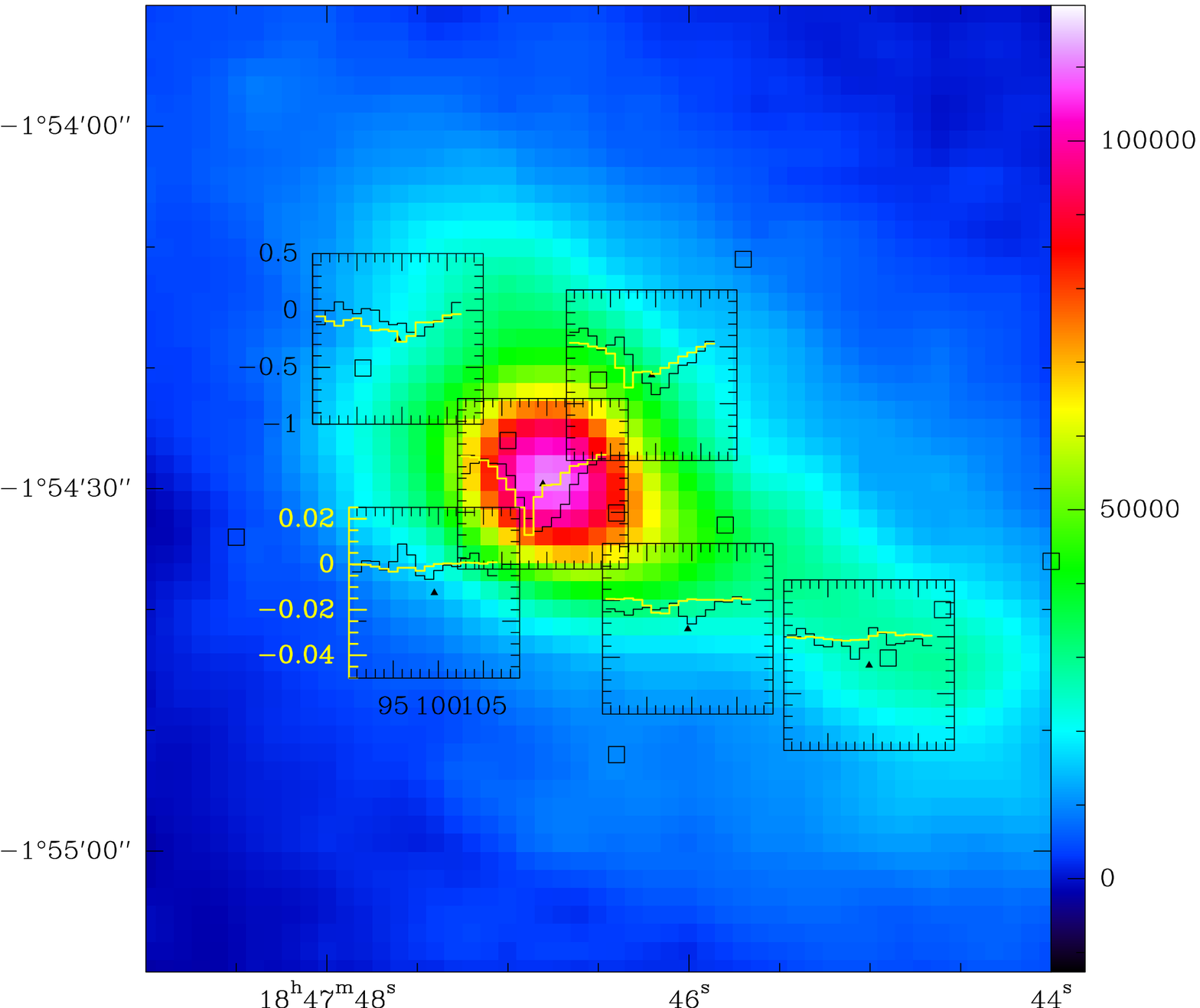}
	\caption{ Observed \hoI\ spectra (black line, temperature scale in black) and predicted \hoI\ (yellow line, yellow temperature scale) due exclusively to the cool outer material which is believed to be responsible for the H$_2^{16}$O absorption. The zero levels are at the same position but there is a factor 4 difference in the scales, showing clearly that the absorption from the cool material is far from sufficient to explain the observed absorption.
Units are $\kms$ for the velocity and Kelvin for the line temperature.  The background is a SCUBA \SI{450}{\micro\metre} image with a linear scale and a peak at 110 Jy/beam.  The triangles indicate the positions of the spectra. The open squares indicate the Louvet et al 2014 sources.}
	\label{figs_Fig11}
\end{figure*}

\section{Velocity structure of the W43 core}
\label{sec:cloud_velocities}

The comparison of the absorptions of the main H$_2$O (\hoK\ and \hoC) water lines and of the H$_2^{18}$O (\hoI\ and \hoA) lines (see Fig. \ref{figs_Fig02}, Fig. \ref{figs_Fig03}, and Fig. \ref{figs_Fig09}) shows that, at all the detected positions, the blue side wing of the main water line has no H$_2^{18}$O counterpart. Hence, we think that this blue side of the main water absorption is due to a cold foreground material, possibly surrounding the MM1 core.
Though an outflow absorption is partly taken into account in the models in Section \ref{sec:warmproto}, we cannot rule out a low density outflow as the source of this absorption.
Fig. \ref{figs_Fig05b} also shows that the redshifted material is located at the immediate north-east off the MM1 core and that the blueshifted material is located south-west of MM1. The central and deepest part of the absorption covers both regions. Though the H$_2^{18}$O absorption probably comes from the inner region of the source, it is also seen off core towards the south-west MM1 following the extended continuum emission. 


\subsection{Central velocity gradient}

Due to its lower optical depths, \hoI\  traces the inner material whereas the main isotope is optically thick in the surrounding cool gas, such that it does not trace the inner material.
Figures \ref{figs_Fig06} and \ref{figs_Fig07} show respectively the main isotope and the \hoI\ ground state absorption lines at selected positions around MM1. 
Figure \ref{figs_Fig06} compares the velocity profiles of \hoK\ spectra between two positions,  chosen symmetrically relative to MM1 and separated by more than a full beamwidth, to the SW and NE of MM1. These positions were chosen to be along the extended continuum object, far enough not to detect the emission from the central source but in a region where the absorption is still clearly detected.
It is immediately apparent that the W43 MM1 absorption shows a velocity gradient whereas the \CFC\ at $80\,\kms$ does not. 
The \CFC s absorptions are rather identical in \AF\ where the main absorption shows a kind of complementary profiles. The blue wing is in emission for one position and in absorption for the other one. The opposite behaviour is seen for the red wing. This inversion occurs, in velocity, symmetrically to the $\simeq 99\kms$ source velocity.

The \hoI\ is not detectable as far out but Fig. \ref{figs_Fig07} shows a similar velocity gradient in the same direction as for the envelope. It shows four \hoI\ spectra at positions surrounding MM1 revealing a velocity gradient in the \hoI\ ranging from $\simeq 98 \kms$ in the south-west of MM1 to  $\simeq 102 \kms$ in the north-east part of the MM1 core. 
The vertical line at the central velocity makes it clear that there is a SW-NE gradient but not SE-NW.  

The \CO\  emission line (Fig. \ref{figs_Fig02}, Fig. \ref{figs_Fig07}), whose energy is above $200\,\K$, comes from the warm central regions. It has the same gradient but the \hoI\ absorption is consistently redshifted with respect to the \CO, showing that the infall onto the main source is visible throughout the inner region.
As for the \hoI\ line, the \CO\ line velocity shift of $0.7 \kms$ is seen from south-west (offset -8\arcsec,-8\arcsec) to north-east (offset +8\arcsec,+8\arcsec), going from 97.9 to $98.6 \kms$. While the size of our beam and the signal-to-noise level of the observations does not allow us to measure the size of the region from which \CO\ emission is detected, the shift shows that it is not point-like even to our $20\,\arcsec$ beam. The small velocity shift shows that the core is not rotationally supported, in agreement with the presence of the infall signature.


\subsection{\hoI\ versus \hoE }
\label{sub:h218o_versus_h2o_987}

Figure \ref{figs_Fig10} compares the \hoI\ profile to the \hoE\ absorption profile at the MM1 position. We first fit a Gaussian profile to model the wide component (see lower left spectrum in figure). Once done, we removed it from the spectrum. Then, we fit the emission part of the remaining spectrum by another Gaussian profile (see lower right spectrum in figure) that we again remove. The hypothesis here is that both the wide and the narrower emission parts of the line profile can be modeled by Gaussian profiles. The last spectrum fits then the absorbing component in the \hoI\ line. We compare its profile to the \hoE line by scaling both lines to the same intensities (see upper box in figure). Both profiles are almost identical in width. The little dip in the red wing is even the same in the two spectra.  We conclude that both lines originate from the same material.

\subsection{The  wide component high-velocity gas}
\label{sub:the_wide_component}

A wide velocity component is observed in several main water lines \citep[see ][]{Herpin-W43-2012}. We detect  it in our maps of  \hoC\ , \hoE\ and \hoK.
The wide component in the \hoC\ map is detected at all observed positions but not used in our discussion because of the large beam size at 557\,GHz.
Because of a poor rms and of some baseline problems there are large uncertainties 
for the \hoE\ spectra. Still, it is possible to map  the wide $\simeq 40\kms $ component wings of \hoE\ as seen in Fig. \ref{figs_Fig08} where 
one sees the red and blue wings bracketing the central absorption in the southern part of MM1.\\
We searched for traces of systematic motion in the high-velocity gas seen in \hoI\ emission. 
Fig. \ref{figs_Fig08} shows the spatial distribution of the line emission, with red-shifted gas in red, blue-shifted gas in blue, and intermediate velocity gas in green. 
The quiescent gas is seen to trace a flattened envelope which is also seen in the continuum (Fig. \ref{figs_Fig01}), while the high-velocity gas is seen to trace two lobes which fall on a line which is approximately perpendicular to this envelope.
If these lobes are due to a bipolar outflow, Fig. \ref{figs_Fig08} indicates a size of $\sim$36'' end-to-end or $\sim$18'' peak-to-peak, which at 5.5 kpc \citep{2014ApJ...781...89Z} corresponds to 0.48-0.96 pc. 
While the map data in Fig. \ref{figs_Fig08} indicate a velocity range of only $\sim 31 \kms$ FWZI or $\sim 25 \kms$ FWHM for the outflow, the pointed spectrum in Fig. \ref{figs_Fig10} shows that \hoI\  emission is detected at higher velocities, up to $\sim 50 \kms$ FWZI.
Furthermore, Fig. \ref{figs_Fig08} indicates a low degree of collimation for the outflowing gas, as often seen in single-dish observations of outflows from massive star-forming regions \citep{2005ccsf.conf..105B}.

Outflow activity from W43-MM1 has been observed before by \citet{2014ApJ...783L..31S}, who used the SMA to image the CO 3-2 line emission from this region at 5'' resolution.
Their data reveal gas shifted to velocities up to $\sim 70 \kms$, which is spatially resolved into 3 distinct bipolar outflows, which together cover an area of 20-30'' in size. 
Thus, our observed high-velocity \hoI\ emission may trace dense gas associated with the high-velocity CO seen with the SMA, and the apparent low collimation may be due to confusion between multiple bipolar outflows.

\section{ Envelope absorption and velocity gradient}

Figure \ref{figs_Fig09} points to the same effect: the blue wing is seen in absorption to the SW and the red wing to the NE.  The total velocity shift is around 4 to 5 $\kms$

A major question is how one can directly observe the massive protostellar core.  It appears that the main isotope \hoK\  absorption at 1113\,GHz comes from the envelope rather than the region above 100\K\ \citep{Herpin-W43-2012}.
In order to assess whether the \hoI\ absorption is due to the cool envelope or the protostellar core, we used the line to continuum ratio of the \hoK\  line to estimate the optical depth at 1113\,GHz, implicitly assuming that all of the absorption is due to the envelope.
With a minor calibration error, one could easily underestimate or overestimate the \hoK\ optical depth.  If the absorption were underestimated, one would expect to have extreme variations in the optical depth as e.g. a 99.99\% absorption will yield a much higher optical depth than 99\% absorption and the variations due to noise are at this level.  If we have overestimated the absorption, then the \hoI\ absorption will also be overestimated.  The absorption is shown in Fig. \ref{figs_Fig05b}, where the highest contour is at an absorption level of 78\% over a width of $4 \kms$, consistent with the somewhat higher absorptions we obtain for the $2 \kms$ width used in these calculations.
Then, assuming an abundance ratio H$_2$O/H$_2^{18}$O of 322, corresponding to the galactocentric radius of W43 \citep{1994ARA&A..32..191W}, we calculate the optical depth and (using the continuum flux) expected absorption line intensity.  This can be directly compared to the observed \hoI\ absorption spectra.

Figure \ref{figs_Fig11} shows the observed  \hoI\ absorption and predicted \hoI\ envelope absorption spectra superposed on a SCUBA \SI{450}{\micro\metre} map.  It can be seen that while there is \hoI\ absorption associated with the envelope, the expected line intensity is below 0.04\K\ everywhere.  The observed \hoI\ absorption reaches 1\K\ in the central regions, incompatible with the hypothesis that the \hoI\ absorption is due to the envelope.  In both the predicted envelope spectra and observed spectra, \hoI\ lines are predicted/detected in regions along the major axis of W43 MM1.  

Therefore, the \hoI\  absorption is due to the central core region which has $T_{mb}>100\K$.  As also shown in Figure \ref{figs_Fig09}, the velocities of the \hoK\ and \hoI\ lines are not the same, further illustrating that they trace different components. The \CO\ emission (Fig. \ref{figs_Fig02}), tracing warm dense material and observed at the same positions with exactly the same beam size, is systematically at a less redshifted velocity than the \hoI\, implying that the water over 100\K\ is infalling.  

Figure \ref{figs_Fig05} and Fig. \ref{figs_Fig05b} visualize the absorbed fraction, for various velocity ranges, of the \hoK\ line.  The situation in Fig. \ref{figs_Fig05b} is clearly different compared to Fig. \ref{figs_Fig05}. The \CFC\ mapped in Fig. \ref{figs_Fig05} do not show the same localized structure seen in Fig. \ref{figs_Fig05b}. The three upper panels of Fig. \ref{figs_Fig05b} show respectively the blue wing, the central part where the absorption is generally higher, and the red wing.  There is little absorption to the NE in the blue wing and to the SW in the red, showing a velocity gradient in the envelope.

\hoI\ emission/absorption is also present outside the central beam.  We believe that this is due to the other massive dense cores found by Louvet et al. in W43 MM1.  Their positions are indicated on Figure \ref{figs_Fig11} and this can be compared with the \hoI\ optical depth as seen in Figure \ref{figs_Fig09}, showing the same result in a slightly different manner.

\begin{figure}
	\centering
	\includegraphics[keepaspectratio,width=8.7cm]{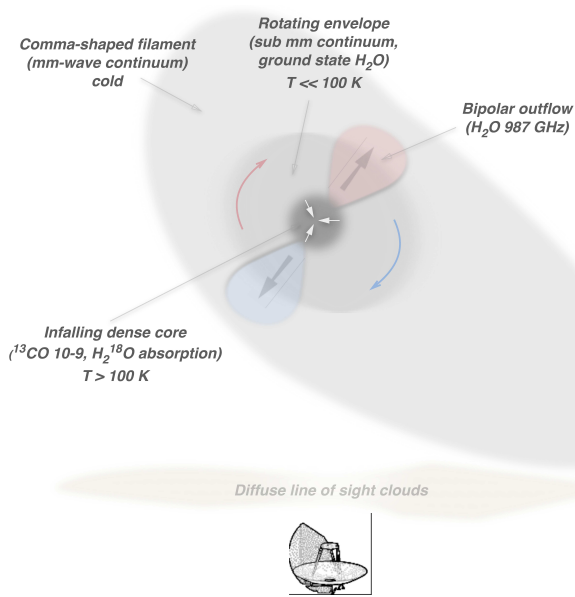}
	\caption{ Drawing of the various W43 MM1 components derived from the spectral maps. Arrows sketch the corresponding velocity fields (infall, outflow and rotation). The observing direction is indicated by the bottom telescope image whose dimension give an idea of the beam size.}
	\label{sketch}
\end{figure}
\section{Source structure and Infall rate}

Based on results of the previous sections, Fig. \ref{sketch} sketches out the kinematics and physical structure of W43 MM1, sizes and geometry of the various components are roughly indicative.
The extended comma shape continuum component contains the infalling W43 MM1 core embedded in a colder envelope. The saturated \hoK\ water line comes from the colder envelope whereas \hoI\ and \CO\ come from the hotter infalling core. The envelope component is resolved by the Herschel beam at $1113\,GHz$. Its possible rotation is suggested by the water lines velocity gradient seen around the core along the comma shape axis. This sketch does not show the many cores sub structure seen in \cite{Louvet-2014-w43mm1} and, gives only the global orientation of the outflow seen in  \hoE.

\subsection{The cold foreground material}

\begin{figure}
\centering
\includegraphics[width=8.7cm]{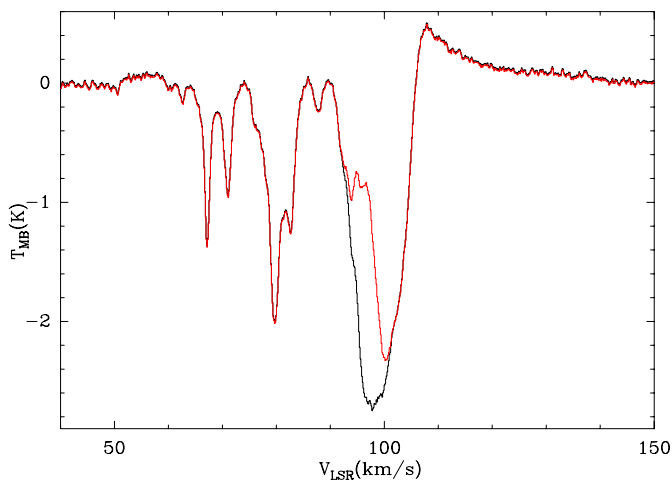}
\caption{Original HIFI spectra of the \hoK line in black (from \citet{Herpin-W43-2012}). In red the spectra after removing of the cold component.}
\label{1113_compare}
\end{figure}

As explained in Sect. 4 and Fig. \ref{figs_Fig09}, a cold foreground cloud embedding MM1 contributes to the blue side of the line profile of the para ground state line of water. We modeled this blending component using our map data as follows. Figure \ref{figs_Fig05} (see left box) shows that the opacity of this material can, at first, be considered rather constant in between MM1 and the south-west part of the map allowing us to make the raw hypothesis that the physical conditions of this material are similar enough both at MM1 and at a location around the offset (-21\arcsec,-14\arcsec). We fit the line at this off core position and, we then scale this result using the continuum ratio at MM1 and at the off core position in order to remove it from the  spectrum observed towards MM1.  
Subtracting this absorption from the pointed spectra of MM1 from  \citet{Herpin-W43-2012}, we get an estimate of the {\em true} water line profile toward the massive dense core MM1 (see Fig. \ref{1113_compare}). The new line profile is of course narrower and the absorption less deep, now rather peaking at $100 \kms$, i.e. red-shifted relative to the source velocity. 

\begin{figure}
\centering
\includegraphics[width=8.7cm]{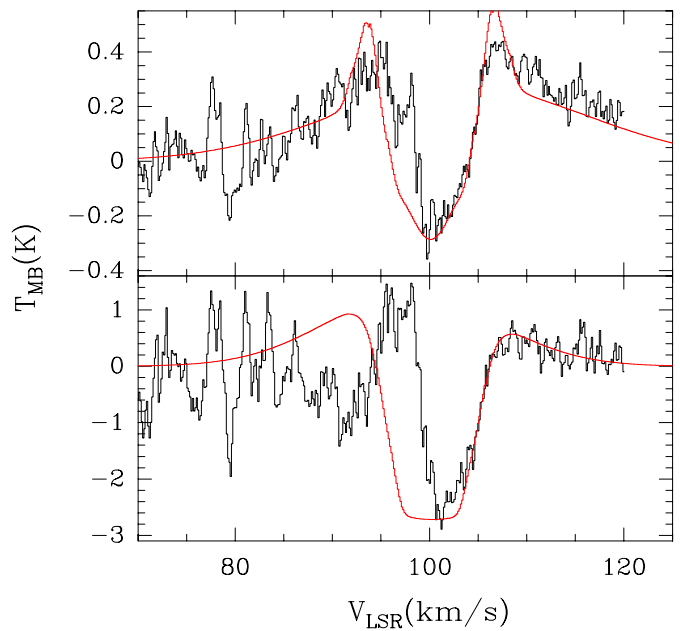}
\caption{HIFI spectra after removing of the cold component for the 557 (top) and 1669\,GHz (bottom) lines. In red is shown the new model.}
\label{1669_557_compare}
\end{figure}

This blue side cold component is also seen in the ortho ground state lines \hoC\ and \hoN. Unfortunately, missing the corresponding H$_2$O line maps, we cannot apply the method described above for the \hoK\ line profile. Still, we can get an estimate of this blue side blending by comparing the many \CFC s components seen in \hoC\ and \hoN\ between 0 and 90 $\kms$ to their \hoK\ counterparts. We note that their integrated intensity (for both \hoC\ and \hoN) can be scaled with almost the same value at all velocities from each \hoK\ counterpart. From our hypothesis that the \hoK\ blue side material 
has a \CFC\ nature 
we can then model it for the \hoC\ and \hoN\ main water line by scaling our fit for \hoK\ with the scale ratio found for the \hoC\ and \hoN "cold" components.
We can now derive the corrected \hoC\ and \hoN\ water line profiles (see Fig. \ref{1669_557_compare}).

\subsection{The warm protostellar envelope}
\label{sec:warmproto}
\begin{figure}
\centering
\includegraphics[width=8.7cm]{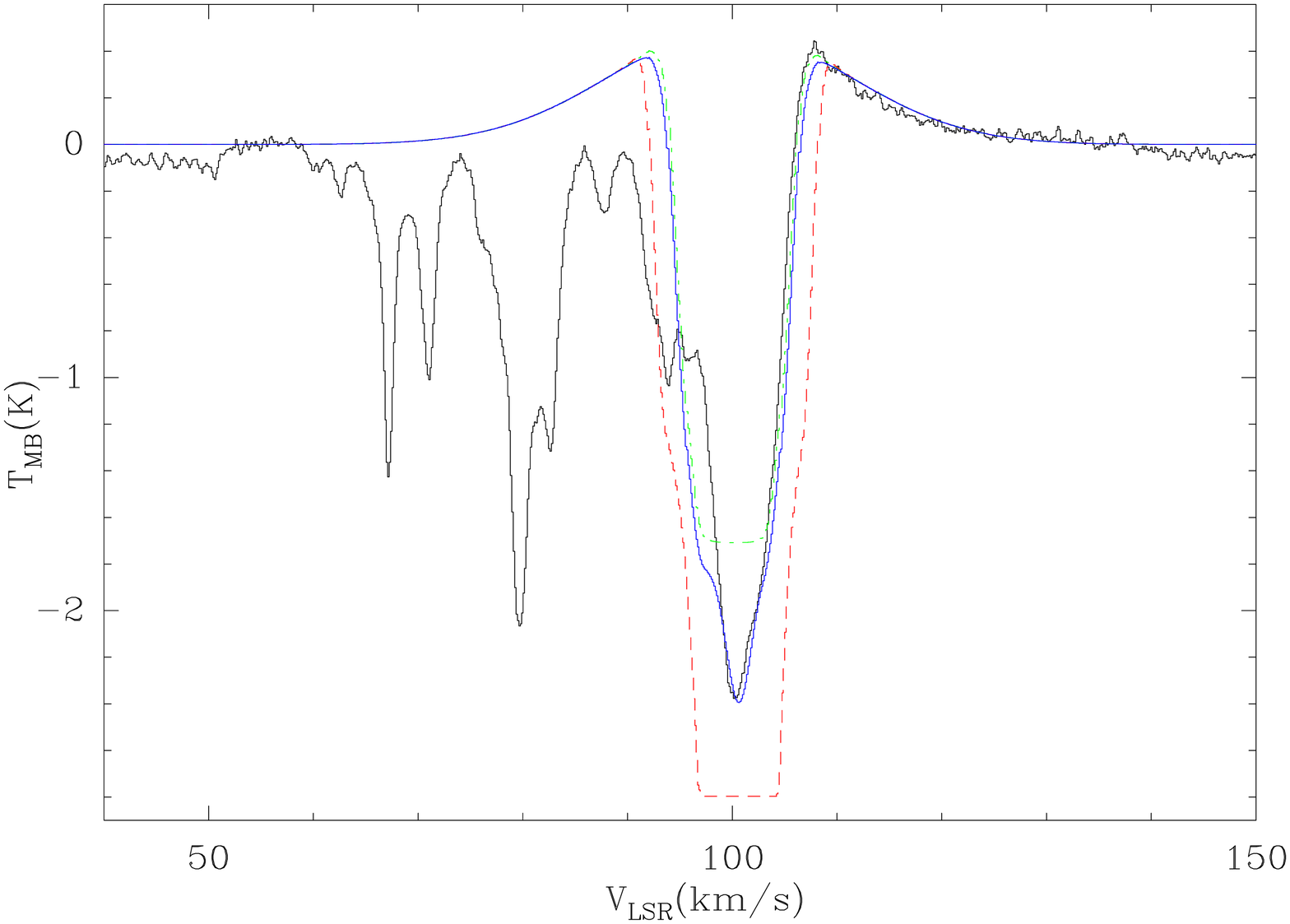}
\caption{New HIFI spectra of the \hoK line in black (from  \citet{Herpin-W43-2012}. The outputs from the model with  \citet{Herpin-W43-2012} parameters, new parameters without, and with a foreground cloud are respectively shown in red, green, and blue.}
\label{1113_compare_model}
\end{figure}

Considering that these new line profiles give us a more realistic view of the MM1 source, we decide to revisit the analysis of  \citet{Herpin-W43-2012} who needed a cold water cloud to properly model the various line profiles. We use the same physical structure for the W43 MM1 source and first apply the same model (same foreground clouds, turbulent velocity, infall and water abundances) than  \citet{Herpin-W43-2012}. Fig. \ref{1113_compare_model} shows that the line profile is not well reproduced: the modeled spectra is too broad and exhibits a saturated deep absorption. The best model is obtained by only decreasing the global outer water abundance from 8 $10^{-8}$ to 8 $10^{-9}$; the inner water abundance and the turbulence velocity increasing with radius have not changed. Surprisingly, a weaker cold component (with an opacity of 0.3) is still needed to reach the absorption dip. The same conclusions apply to the \hoC\ and \hoN\ lines (opacity of 0.4 for the cold component) (see Fig. \ref{1669_557_compare}). 

{Though the outer water abundance in other mid-IR quiet high-mass protostellar objects (HMPO) has been found by \citet{herpin2016}, and also by \citet{marseille2010}, to be of a few $10^{-8}$ (but $1.4\times10^{-7}$ for DR21(OH)), other estimates in some more evolved massive objects lead to lower values: $2.4-6.3\times10^{-9}$ in AFGL2591 \citep[][]{choi2015} and $1-5\times10^{-9}$ for some bright HMPOs, hot molecular cores, and Ultra Compact HII regions \cite{ChoiPhD2015}. In addition, low-mass protostellar objects tend to have an outer abundance of $\sim 10^{-8}$ \citep[][]{vandishoeck2011} or even lower \citep[$3\times 10^{-10}$,][]{mottram2013}. Moreover, \citet{snell2000} or \citet{caselli2012} have estimated very low abundances ($10^{-10}-10^{-8}$) in cold regions. Hence the value derived in our study is compatible with what is found in cold outer regions. In the cold envelopes, water is mainly found as ice and can sublimate through FUV photodesorption \citep[photons from the interstellar radiation field, ISRF, and from cosmic rays interacting with H$_2$, see][]{schmalzl2014}.
Of course, part of this water vapour can then be photodissociated through FUV photons. Variations of the ISRF (e.g. proximity of a nearby bright star) or self-shielding for instance can lead to different outer abundances for different objects.

While we find lower outer water abundances than derived from pointed observations, it is unlikely that this can be generalized to other sources. W43 MM1 is indeed part of a complex cluster embedded in an elongated structure \citep[][]{Louvet-2014-w43mm1}. Only further studies of similar maps observed toward these objects can help to confirm or revise previous water abundances.}

In the new models, as for ${\rm NH}_3\ 3_{2+}\,-\,2_{2-}$ in \citet{2012A&A...542L..15W}, the lines are now centered at $100 \kms$ rather than $98.6 \kms$ before in  \citet{Herpin-W43-2012}, hence red-shifted relative to the source velocity, which is indicative of infall. While in \citet{Herpin-W43-2012}, the ground-state lines were dominated by the cold cloud (in absorption), and thus were not showing any infall signature, these new water line profiles do indicate infall. This infall signature is now well reproduced by this new model using the same infall velocity ($-2.9 \kms$) than \citet{Herpin-W43-2012}, then mainly derived from the 987 and 752\,GHz water lines. 
Before subtraction of the widespread cool component, the low-energy H$_2$O lines were not sensitive to infall, which could only be traced at higher energies and/or with isotopologues.  Now the H$_2$O lines yield the same level of infall.
The main consequence of this new outer water abundance is that the water content in the outer region is an order of magnitude lower than previously estimated: 3.7 $10^{-4}$ M$_{\odot}$ (instead of 3 $10^{-3}$ M$_{\odot}$).  

\section{Conclusions}

The main conclusion from this work is that map data are an essential complement to pointed observations to ensure reliable estimation of physical and chemical parameters from sub-millimeter line spectra.
In particular, molecular abundances derived only from pointed spectra may be off by a factor of $\simeq 10$, while estimates of turbulent velocities and infall / outflow rates are less affected.

Our second conclusion is that lines of H$_2^{18}$O, as well as excited-state H$_2$O lines, trace the inner envelopes of high-mass protostars, where abundances are enhanced due to grain mantle evaporation and/or high-temperature gas-phase chemistry.
Deriving these abundances requires careful subtraction of the contribution from the outer envelope, however.

Third, we find that warm chemically enriched gas in W43 is not confined to the MM1 position, but is also found towards other cores seen in warm dust emission by  \citet{Louvet-2014-w43mm1}.

Finally, the HIFI maps of the W43 region reveal velocity gradients in the envelope of the MM1 core as well as in the surrounding gas, which are indicative of rotation. Figure \ref{sketch} summarizes the global picture derived for the MM1 region.

\bibliographystyle{aa} 
\bibliography{biblio}{} 

\end{document}